\documentclass[floatfix,aps,pre,showpacs]{revtex4}

\usepackage{graphicx}
\usepackage{amsmath}
\begin{document}

\title{Inference algorithms for gene networks:
A statistical-mechanics analysis}

\author{A Braunstein$^{1,2}$, A Pagnani$^1$, M Weigt$^1$ and R
Zecchina$^{1,2}$}

\affiliation{$^1$ Institute for Scientific Interchange, Viale S. Severo
65, I-10133 Torino, Italy} 

\affiliation{$^2$ Politecnico di Torino, Corso Duca degli Abruzzi 24,
I-10129 Torino, Italy}





\date{\today}

\begin{abstract}
The inference of gene-regulatory networks from high-throughput
gene-expression data is one of the major challenges in systems
biology. This paper aims at analysing and comparing two different
algorithmic approaches. The first approach uses pairwise correlations
between regulated and regulating genes, the second one uses
message-passing techniques for inferring activating and inhibiting
regulatory interactions. The performance of these two algorithms can
be analysed theoretically on well-defined test sets, using tools from
the statistical physics of disordered systems like the replica method.
We find that the second algorithm outperforms the first one since it
takes into account collective effects of multiple regulators.
\end{abstract}


\maketitle

\section{Introduction}

\begin{figure}[htb]
\vspace{0.2cm}
\begin{center}
\includegraphics[width=0.45\columnwidth]{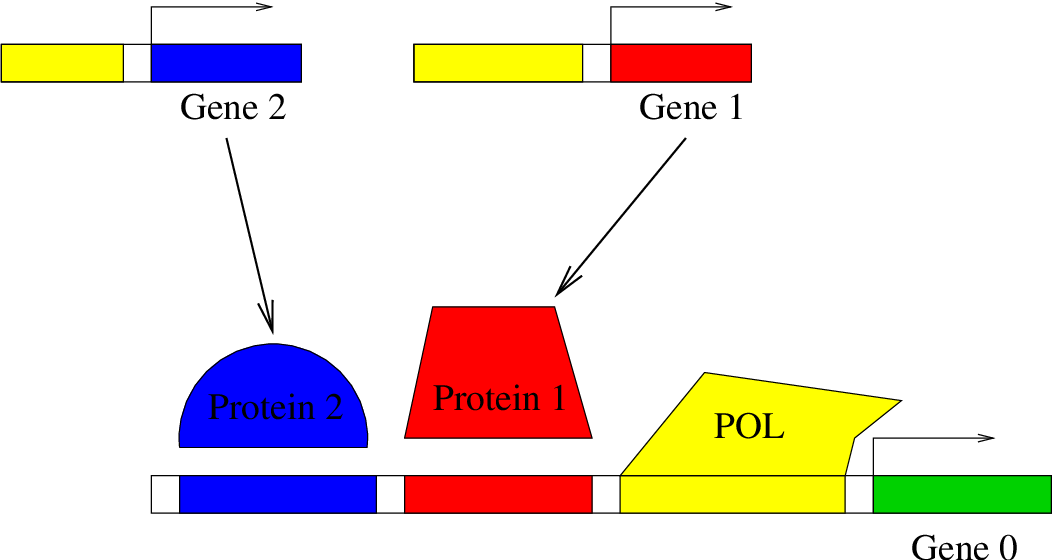}
\end{center}
\caption{Cartoon of transcriptional regulation of gene 0 by two
the transcription factors protein 1 and protein 2.}
\label{fig:gene_regulation}
\end{figure}

According to the central dogma of molecular biology, there is a
directed information flux from genes via mRNA to proteins whose
aminoacid sequence is coded in the genes. This simplified view leads,
however, to a paradoxical situation: All cells in a multicellular
organism carry the same genetic information, {\it i.e.}, the same DNA
sequence, but they differ largely in their protein content which is
responsible for the different functionality of, {\it e.g.}, neurons,
liver cells, skin cells etc. The crucial process which allows for cell
differentiation is {\it gene regulation} \cite{Alberts}, as sketched
in Fig.~\ref{fig:gene_regulation}: The transcription of a gene to mRNA
is achieved by RNA polymerase (POL) or, in higher organisms, a protein
complex containing polymerase. The binding of this polymerase to the
DNA in the so-called promoter site depends, however, on the presence
or absence of other proteins, the transcription factors, which have
binding sites close to the promoter site. These transcription factors
are themselves proteins, so they are coded for in genes in other
regions of the DNA. We say that there is a {\it gene-regulatory
interaction} from genes 1 and 2 to gene 0. Since the expression of
genes 1 and 2 may be regulated by further transcription factors, the
set of all such interaction forms a complex {\it gene-regulatory
network} (GRN).

If we knew the gene-regulatory network of an organism, it would be an
interesting question to infer the dynamical behaviour and stationary
points of these networks to understand the process of cell development
and differentiation. In addition, this knowledge would allow us to
numerically predict the outcome of genetic interventions, and
eventually to identify new drug targets for medical treatment. It is,
however, a very complicated, time-consuming and cost-intensive task to
experimentally determine the detailed structure of such networks
using techniques like gene-knockout, chromatin immuno-precipitation
(ChIP) etc.  In fact, genome-wide networks are only know for simple
organisms like {\it E. coli} \cite{Alon1} and baker's yeast
\cite{Kepes,Alon2}, whereas for higher organisms only few functional
modules are known, see {\it e.g.} \cite{Davidson,Albert}.

On the other hand, today it is relatively easy to monitor the gene
expression on a genome-wide scale, the most important experimental
tools here are microarrays (DNA-chips) which measure the simultaneous
abundance of tens of thousands of mRNAs, {\it i.e.} of the products of
the transcription process, which later on get translated into
proteins.  It seems therefore an enormously interesting task to infer
gene-regulatory interactions from expression data using
bio-informatics means. Even if such an approach cannot be expected to
replace a direct experimental determination of such interactions, it
may well serve as a guide to design efficient and focused
experiments.

In the literature, various inference algorithms are discussed. A first
class deals with relevance networks, measuring the correlation
resp. mutual information of the expression of gene pairs
\cite{RN1,RN2}. This approach obviously may include also indirect
interactions, since also second or third neighbours in a regulatory
network may be correlated. This is taken into account in the algorithm
ARACNe \cite{Aracne1,Aracne2}, which uses information-theoretic
arguments to prune all those links which can be explained by indirect
interactions. In Bayesian and dynamic Bayesian networks
\cite{Bayes1,Bayes2,Bayes3,Bayes4}, also the directed
structure of causal dependencies between gene expression levels is
inferred: A global scoring function is assigned to arbitrary networks,
and the highest scoring network is considered to be the best possible
inferred network. Unfortunately, finding this network is an NP-hard
task, it can reasonably accessed only using heuristic tools (greedy
algorithms with restarts, simulated annealing...) which are likely to
get trapped in local extrema of the scoring function. A somewhat
similar approach are probabilistic Boolean networks \cite{PBN1,PBN2}
which exhaustively scan all gene pairs or triplets in order to see in
how far they determine the expression level of an output gene. Even if
being polynomial, this algorithm is restricted to relatively few
genes and regulatory functions with at most three inputs.

The inference of gene-regulatory networks is unfortunately plagued by 
serious problems based on the quantity and the quality of available data. 
Some of the most relevant limitations are listed below:
\begin{itemize}
\item[{\it (i)}] The number $M$ of available expression patterns is in 
general considerably smaller than the number $N$ of measured genes.
\item[{\it (ii)}] The available information is, in general,
incomplete.  The expression of some relevant genes may not be
recorded, and external conditions corresponding, {\it e.g.}, to
nutrient, mineral, and thermal conditions are not given. Micro-arrays
measure the abundance of mRNA, whereas gene regulation works via the
binding of the corresponding proteins (transcription factors) to the
regulatory regions on the DNA. Protein and RNA concentrations are,
however, not in a simple one-to-one correspondence.
\item[{\it (iii)}] The data are noisy. This concerns biological
noise due to the stochastic nature of underlying molecular processes, as
well as the considerable experimental noise existing in current 
high-throughput techniques.
\item[{\it (iv)}] Micro-arrays do not measure the expression profiles of 
single cells, but of bunches of similar cells. This averaging procedure 
may hide the precise character of the regulatory processes taking place in
the single cells.
\item[{\it (v)}] Non-transcriptional control mechanisms (chromatin 
remodelling, small RNAs etc.) cannot be taken into account in expression
based algorithms.
\item[{\it (vi)}] A last problem is a conceptual one: Searching for
dependencies between the expression levels of genes does not
automatically imply their direct physical interaction in
transcriptional regulation. To give an example, strongly coregulated
genes may appear as good predictors of each other, without having any
direct interaction.
\end{itemize}                             
The listed points obviously lead to a limited predictability of even the most
sophisticated algorithms. It is therefore very important to test various
algorithms on the basis of well described data sets containing some or all 
of the before-mentioned problems: Only a critical discussion on artificial
data sets allows for a sensible interpretation of the algorithmic outcomes
when run on biological data.

In this paper, we present a recently introduced statistical-physics
motivated inference algorithm build on a message-passing technique
which, in a sense, is equivalent to the Bethe-Peierls approximation of
statistical physics \cite{BP-infer,Japan}. We further use
statistical-physics tools, more precisely the replica method developed
in the theory of disordered systems and spin glasses, to characterise
the performance of this algorithm on artificially generated data, and
to compare it to relevance resp. ARACNe networks. We will show that
the message-passing technique is able to take into account also
collective effects in the common action of various transcription
factors, and therefore it outperforms local pair-correlation based
algorithms.

The structure of the paper is as follows: In the following section, we
formalise the network inference problem. In the third section, we
review two types of inference algorithms, namely pair-correlation
based methods and a message-passing technique. Than we introduce
a generator for artificial data and perform first numerical
tests. Sec.~\ref{sec:analysis} contains a statistical-mechanics
approach to the performance of both algorithms, and a careful analysis
of their behaviour under various (artificial data) conditions. At the
end, we conclude this work, and discuss extensions of the message
passing techniques which are necessary for a successful application
to biological data.

\section{The network inference problem}

\begin{figure}[htb]
\vspace{0.2cm}
\begin{center}
\includegraphics[width=0.55\columnwidth]{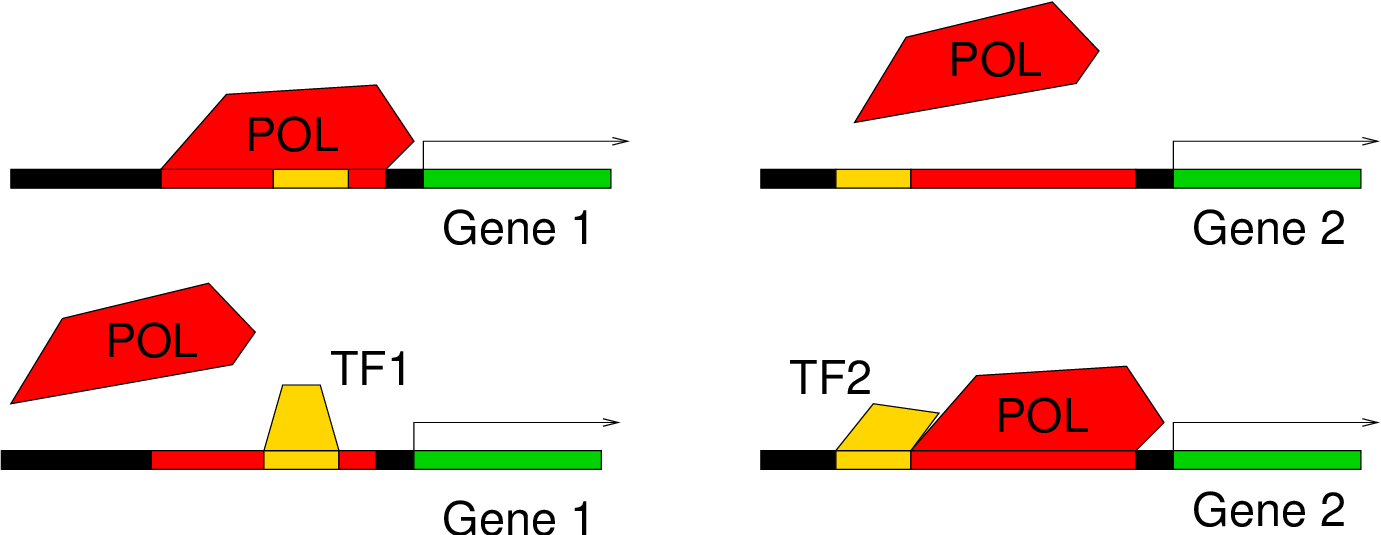}
\end{center}
\caption{Schematic representation of transcriptional repression (left)
and activation (right). Repression works via the competitive binding
of POL and the TF to overlapping binding sites, wheres activation uses
a short-range attraction between POL and TF when attached to close-by
binding sites.}
\label{fig:repressor-activator}
\end{figure}

Let us first formalise the network-inference problem. Given are
stationary points of the regulatory network, or gene-expression
profiles (Note: Temporary expression data can be learnt with a modification
of the presented algorithm. Here we concentrate fully on static data.) 
They can be written in the following form
\begin{equation}
x_i^\mu,\hspace{1cm}i=0,...,N,\ \ \mu=1,...,M
\end{equation}
with $i$ enumerating the considered genes, and $\mu$ the different
expression patterns (profiles). In this work, we assume the expression
values to be binarized, $x_i^\mu=-1$ signifies that gene $i$ has very
low expression in patterns $\mu$, whereas a high expression is denoted
by $x_i^\mu=1$. Note that we use an Ising spin notation instead of a
more standard Boolean one, but this choice will render the technical
presentation better accessible. Such a binary representation seems to
be oversimplified since gene expression is characterised by mRNA
concentrations, but it is applied successfully in various biological
examples \cite{Davidson,Albert,PBN2}.  It is expected to work pretty
well since gene-regulatory functions show frequently a switch like
behaviour \cite{Hwa}. The simplest examples, repression and
activation by a single TF, are illustrated in
Fig.~\ref{fig:repressor-activator}.

At this level, the inference problem can be considered to factorise over
the output variables: We can first reconstruct the regulators for variable
0, then for variable 1 etc. until we reach variable $N$. In our 
presentation we concentrate therefore to one of these subproblems:
Without loss of generality, we are interested in the question in how far 
and in which way gene 0 is regulated by the genes 1,...,N. We can distinguish 
two subsequent subproblems
\begin{itemize}
\item {\it Topology}: Which genes do influence gene 0, and which other genes
do not influence gene 0. This result of this question are directed
links from the regulating genes to the regulated gene 0.
\item {\it Functionality}: How do these genes influence the expression of
0? Here we want to restrict this question to the simplest types of
interactions, classifying the input genes 1,...,N as {\it activators},
{\it repressors} or irrelevant for gene 0.
\end{itemize}
Note that the factorised approach may be oversimplified if experimental noise 
is considered. Here we are, however, not interested in giving a biologically
realistic modelling, but in a theoretical analysis of the proposed message-passing
algorithm. We therefore keep this assumption since it allows a more elegant
and clear presentation of the results.

As mentioned, we aim at classifying the directed
interactions from gene $i$ to gene 0 by a ternary variable,
\begin{equation}
J_i = \left\{
\begin{array}{rl}
-1 & {\mbox{ if gene $i$ represses the expression of gene 0}}\\
0 & {\mbox{ if gene $i$ does not regulate gene 0}}\\
1 & {\mbox{ if gene $i$ activates the expression of gene 0}}
\end{array}
\right.
\end{equation}
Again, this classification is oversimplified with respect to
biological reality. First, it does not include the fact that there may
be strong and weak activators or inhibitors, all of them are just
characterised by one single variable. Second, there may be
interactions where a gene changes its activating or inhibiting
character in the presence of other genes. To give an example, the XOR
function of two variables cannot be represented within the
above-mentioned classification. 

It would be possible to overcome this simplification by introducing
more complicated variables. On the other hand, this would increase the
number and complexity of the model-parameters, and consequently the risk
of overfitting. The task of extracting a good predictor from few noisy
data requires the model to depend on as few paramters as possible.

To infer this interaction vector, we have also to set up a minimal 
functional model for
the joint interaction of all relevant input genes. Having classified
the input genes by this ternary variable, we can sum up the total
influence of an expression pattern $\vec x= (x_1,...,x_N)$ on gene
0. If it overcomes some threshold $\theta$ (which, for clarity of
notation, we set to zero in the following), the expression
of gene 0 is expected to be enhanced, below this threshold the
expression level of gene 0 is repressed. At the Boolean level, we can
therefore model the regulatory interaction by the threshold function
\begin{equation}
\label{eq:perz}
x_0 = {\rm sign} \left( \sum_{i=1}^N J_i x_i \right)\ .
\end{equation}
Note that, slightly more than one decade ago, this function has found
considerable attention in statistical physics in the context of the
theory of artificial neural networks \cite{HeKrPa,EnBr}, it describes
a diluted perceptron. In fact, the technical approach presented below
consists of an adaption of the well-known Gardner calculus of
perceptron learning \cite{Ga} to the gene-regulatory problem.

Using this model, we are looking for classification vectors $\vec J =
(J_1,...,J_N)$ which are as compatible as possible with the input data
$\vec x^\mu,\ \mu=1,...,M$. We therefore introduce a Hamiltonian
\begin{equation}
\label{eq:hamiltonian}
{\cal H}(\vec J) = \sum_{\mu=1}^M \Theta \left( -x_0^\mu\cdot 
\sum_{i=1}^N J_i x_i^m \right)
\end{equation}
which counts the number of patterns whose output contradicts the model
(\ref{eq:perz}). The $\Theta$ symbolises the step function,
which jumps from zero for negative arguments to one for non-negative
arguments. One aim of the algorithm will be to minimise
this Hamiltonian. 

To get reasonable results, we will also implement another biologically
motivated constraint. Gene regulatory networks are {\it diluted}, i.e. 
the number of transcription factors for one gene is generally restricted
to a small number. In the best known regulatory network, the one of
bakers yeast, one finds evidence that the number of genes regulating
commonly one gene is distributed according to a narrow exponential
distribution, which can be explained by physical constraints for the
positioning of transcription factor binding sites. To take into
account this diluted character, we will also control the number
\begin{equation}
\label{eq:neff}
N_{eff}(\vec J) = \sum_{i=1}^N \left( 1- \delta_{J_i,0} \right)
\end{equation}
of effectively present, {\it i.e.}, non-zero links via a conjugated
external field. This {\it a priori} bias towards diluted graphs helps
again to avoid overfitting of the data, since it reduced the entropy
of the search space.

This idea is closely related to methods of regression and graphical-model
selection using $\ell_1$-parameter regularization discussed in the 
machine-learning literature. The original idea was developped by Tibshirany 
for the case of sparse regression in linear models \cite{tibshirany}, but it
can be extended to more general convex optimization problems \cite{wainwright,
banerjee,koller,schmidt}. In all these cases, the optimization of the cost 
function is restricted by the $\ell_1$-norm $\sum_i |J_i|$ of the model parameters 
$J_i$. The cusp-like structure of this constraint in $J_i=0$ forces a part 
of the parameters to be exactly zero. In the presence of sufficient 
data this method can be shown to asymptotically reproduce the correct 
topological structure of the underlying model \cite{meinshausen}. In principle,
one could use any $\ell_n$-norm $\sum_i |J_i|^n$ for regularizing parameters. 
However, only for $0\leq n\leq 1$ some of the optimal parameters assume values 
equal to zero; for $1\leq n$ the constraint is convex, thus it can be added to 
the cost function without destroying its convexity. Only the $\ell_1$-norm 
fullfils both conditions.
 
Our case is, however, different from these examples. The {\it discrete} nature
of the model parameters renders the model {\it a priori} NP-hard, and convex
optimization cannot be applied. Furthermore, we are interested in ensembles
of possible networks rather than a single network as predicted by convex
optimization.

\section{Inference algorithms}
\label{sec:algorithms}

In this section, we present two classes of inference algorithms. The
first one is based on pairwise correlations between the output
variable $x_0$ and inputs $x_i$. This correlation can be measured in
different ways, two of the most important measures are the connected
correlation coefficient (or its normalised version, the Pearson
correlation coefficient), and the mutual pairwise information of the
two variables. Inference algorithms based on this measure are wide
spread, in their simplest version they result in the so-called
co-expression or relevance network. Such networks normally contain
much more links than the actual regulatory interaction between
transcription factors and regulated genes: Also second neighbours may
still be considerably correlated. This is taken into account in a very
clever way in the algorithm ARACNE which eliminates all those links
which, from an information theoretic point of view, can be explained
via a two-step interaction with an intermediate variable. This
elimination step increases strongly the precision of the algorithm
since it reduces the number of false positives, but it decreases also
its sensitivity cancelling many weak pair interactions.

The second method is the central one of our paper. It is based on a
message-passing algorithm (more precisely a belief-propagation
algorithm) which characterises the statistical properties of the set
of all potential coupling vectors, weighted with respect to
Hamiltonian (\ref{eq:hamiltonian}) and the number of relevant links
(\ref{eq:neff}). We will see that this method is able to - at least
partially - recognise the collective effects between different inputs
in the data-generating rule (\ref{eq:generate}). It therefore goes
beyond simple two-point correlations, at the cost of being related to
a more specific model of gene regulation.

Both algorithms will output a ranking of all potential input variables 
$1,...,N$ according to the considered significance measure, going from the 
most significant genes to the most insignificant
ones. The method itself has no intrinsic criterion where to cut this
list. It his, however, possible to use suitable data randomisation
procedures which give an indication at which level the significance value
of an input may be explained by a random input-output relation (more precisely
one may assign a $p$-value giving the probability that the significance value
or a larger one comes from a random input-output relation).

\subsection{Relevance networks}

Let us first explain pair-correlation based methods, like relevance
networks \cite{RN1,RN2}. They are based on measuring the correlation
between the output variable $0$ and each input variable
$i\in\{1,...,N\}$. The central quantities hereby are the joint
distribution $p(x_0,x_i)$ of the two variables, as estimated from the
data via
\begin{equation}
\label{eq:joint}
p(x_0,x_i) \simeq \frac 1M \sum_{\mu=1}^M \delta(x_0,x_0^\mu)\ 
\delta(x_i,x_i^\mu)\ ,
\end{equation}
and its marginals
\begin{eqnarray}
p(x_0) &=& \sum_{x_i} p(x_0,x_i)\ \simeq\ 
\frac 1M \sum_{\mu=1}^M \delta(x_0,x_0^\mu)
\nonumber\\
p(x_i) &=& \sum_{x_0} p(x_0,x_i)\ \simeq\ 
\frac 1M \sum_{\mu=1}^M \delta(x_i,x_i^\mu)
\ .
\end{eqnarray}
The central question is, in how far the two variables are correlated, i.e. in 
how far the joint distribution is different from the factorised distribution
$p(x_0)p(x_1)$. This can be measured either via the {\it connected correlation
coefficient},
\begin{equation}
\label{eq:corr}
C_{0i} = \sum_{x_0,x_1} \left[ p(x_0,x_i)-p(x_0)p(x_i)\right]\ x_0\ x_i\ ,
\end{equation}
or via the {\it mutual information},
\begin{equation}
\label{eq:mi}
I_{0i} = \sum_{x_0,x_1} p(x_0,x_i) \log\frac{p(x_0,x_i)}{p(x_0)p(x_i)}\ .
\end{equation}
Note that Eq.~(\ref{eq:corr}) is equivalent to the standard definition
of the connected correlation, but the present formulation brings to evidence
the distance measure between the joint and the factorised probabilities.
The mutual information is also known as the Kullback-Leibler divergence
between $p(x_0,x_i)$ and $p(x_0)p(x_i)$. It equals zero if and only if
both are equal, i.e. if $x_0$ and $x_i$ are statistically independent, and 
it is positive otherwise.

These numbers (or, in the case of the correlation coefficient its absolute 
value) can be ordered and give thus a ranking of the inputs according to 
their statistical correlation with the output. One can expect the most
significant correlations to come from direct interactions, whereas weak
correlations are a sign for a non-existing link in the data-generating system
(model or experiment).

Note that this method is not able to detect the {\it directed} character of 
gene regulatory interactions, since the very definition in Eq.~(\ref{eq:corr})
is symmetric against exchange of variables $0$ and $i$. This is a major 
drawback of correlation based methods, and can be eliminated using the 
algorithm proposed in the next subsection.

\subsection{Belief propagation for network inference}
\label{sec:bp}

In this section we present our message-passing algorithm for network
inference.  It has two major differences to the correlation based
method presented before.  On one hand, it is model based
(cf.~Eq.~(\ref{eq:perz})) and therefore may fail in situations where
this model is not sufficiently well-adapted. On the other hand, this
model-based approach allows to take into account the joint action of
all potential regulator variables. Therefore it may well go beyond the
simple two-point correlations used in the previous section.

The basic step is the introduction of a global weight function
\begin{equation}
\label{eq:j-distribution}
W(\vec J\ ) =
\exp\left\{ - \beta {\cal H}( \vec J ) -h N_{eff}(\vec J)  \right\}
\end{equation}
weighing all candidate coupling vectors $\vec J$ ({\it i.e.}, all
potential regulatory networks) with respect to their energy ${\cal H}(
\vec J )$ and the number $N_{eff}(\vec J)$ of non-zero coupling
components, see Eqs.~(\ref{eq:hamiltonian},\ref{eq:neff}) for the
definitions. This weight depends on two not yet specified parameters:
The formal inverse temperature $\beta$ controls the energy. In the
limit $\beta\to\infty$ the weight concentrates in the global minima of
${\cal H}$. The formal external field $h$ controls, on the other hand,
the density of non-zero entries in $\vec J$. Due to the biologically
reasonable request for diluted $\vec J$, we expect this field to
assume high values. There is no obvious strategy of fixing these two
parameters, but we will describe a reasonable heuristic strategy.

Note that the data $\vec x^\mu$ represent the quenched 
disorder, whereas the $N$ components of the vector $\vec J$ are the 
(ternary) degrees of freedom. 
The primary algorithmic aim here is to determine the marginal single-site 
distributions
\begin{equation}
\label{eq:marginal}
P_i(J_i) \propto \sum_{\{J_j\in\{0,\pm 1\};j\neq i\}} W(\vec J)
\end{equation}
where the normalisation constant in this equation is, as usually, given 
by the partition function. A direct calculation of these marginals is a 
computationally intractable task, since Eq.~(\ref{eq:marginal}) contains
the sum over $N-1$ ternary degrees of freedom, i.e. we have to sum up 
$3^{N-1}$ terms. It is therefore necessary to find some - possibly heuristic
- tools which render this task efficiently solvable.

Here we use the idea of message-passing (MP) techniques. These were
initially created for problems with sparse constraints, i.e.,
variables have to be contained in few constraints, and constraints
have to contain few variables.  Typical examples for such problems are
random 3-SAT and colourings or vertex covers on finite-connectivity
random graphs. Here we are in the opposite situation: Each variable
$J_i$ is contained in $M$ constraints given by the expression patterns
$\vec x^\mu, \mu=1,...,M$, and each constraint contains all $N$
variables. Recently the applicability of BP has, however, been
verified in a number of cases ranging from information theory to
perceptron learning \cite{CDMA,UdKa,BrZe}. In \cite{BP-infer,Japan} we
have adopted these techniques to the specific needs of our
network-inference problem.

The BP equations can be easily written down. Each constraint sends a message
to each variable, and each variable sends a message to each constraint:
\begin{eqnarray}
\label{eq:bp1}
P_{i\to \mu} (J_i) & \propto & e^{-h |J_i|}\ \prod_{\nu\neq \mu}
\rho_{\nu\to i} (J_i) \nonumber\\
\rho_{\mu\to i} (J_{i}) & \propto & 
\sum_{\left\{ J_{j},\ j\neq i\right\} } \exp\left\{ -\beta
\Theta\left[ -x_0^\mu\cdot \sum_k J_k x_k^\mu \right]\right\} 
\prod_{j\neq i}P_{j\to \mu}(J_{j})
\end{eqnarray}
The messages have an intuitive interpretation. The variable sends the
information to the constraint: ``If you were not present, I would have
the statistical properties given by $P_{i\to \mu} (J_i)$.'' The constraint,
on the other hand, sends a message to the variable: ``Seen the behaviour 
of the other variables, please behave according to $\rho_{\mu\to i} (J_{i})$
in order to satisfy me.'' Note that there are $2MN$ of these messages,
which form a self-consistently closed set of equations. The true marginal 
distribution can than be estimated from the messages sent by all constraints,
\begin{equation}
P_{i} (J_i) \propto e^{-h |J_i|}\ \prod_{a}\rho_{a\to i}(J_i)\ .
\end{equation}

From a computational point of view, we still have not gained anything. The 
second of Eqs.~(\ref{eq:bp1}) still contains the exponential sum over
all but one $J_j$. This problem can be solved by the following observation:
The dependence on the $J_i$ enters via $\sum_k J_k x_k^\mu$ and via the
factorized distribution $\prod_{j\neq i}P_{j\to \mu}(J_{j})$. For $N\gg 1$
we can apply the central limit theorem and replace the sum over the $N-1$
couplings by a single Gaussian integral, for finite $N$ it will be used as a
computationally efficient approximation. We replace the second line of 
Eqs.~(\ref{eq:bp1}) by 
\begin{equation}
  \label{eq:bp_gauss}
\rho_{\mu\to i} (J_{i})  \propto \int_{-\infty}^\infty \frac{dh}{\sqrt{2\pi}
\Delta_{\mu\to i}} \exp\left\{ -\frac{(h-h_{\mu\to i})^2}{2\Delta_{\mu\to i}^2} 
-\beta \Theta\left[ -x_0^\mu\cdot (J_i x_i^\mu + h )\right]\right\} 
\end{equation}
with 
\begin{eqnarray}
h_{\mu\to i} &=& \sum_{j\neq i} x_j^\mu\ \langle J_j \rangle_{j\to\mu} 
\nonumber\\
\Delta_{\mu\to i}^2 &=& \sum_{j\neq i}\ {x_j^\mu}^2 \left(\ \langle J_j^2 \rangle_{j\to\mu}
-  \langle J_j \rangle_{j\to\mu}^2\ \right)
\end{eqnarray}
where $\langle\cdot\rangle_{j\to\mu}$ denotes the average over
$P_{j\to\mu}(J_j)$. Due to the step-like shape of the Heaviside
function, the right-hand site in Eq.~(\ref{eq:bp_gauss}) can be
expressed via the sum of two error functions, which again leads to a
faster numerical implementation.

Note that, in contrast to what happens in most combinatorial
optimisation problems (and, in particular, also in Bayesian and
dynamic Bayesian network learning), we do not have the aim to
construct one single instance of a ``good'' coupling vector $\vec J$,
because this vector may be quite different from the original
data-generating vector. We are instead interested in characterising
the {\it ensemble} of all sets of vectors. More precisely, the
marginal probability $P_i(J_i)$ gives us a confidence measure for the
hypothesis that the output variable $x_0$ actually depends on $x_i$:
It measures the fraction of all suitable coupling vectors $\vec J$ (as
weighted by Eq.~(\ref{eq:j-distribution})) where the entry $i$ takes
value $J_i$. We can therefore base a global ranking of all potential
couplings in the probability $(1-P_i(J_i=0))$ of having a non-zero
coupling (or alternatively on the average coupling $\langle J_i
\rangle_{i}$).

Having calculated the messages and the marginal distributions, we may
determine the energy of the average coupling vector
\begin{equation}
E = \sum_{\mu=1}^M \Theta \left( -x_0^\mu\cdot 
\sum_{i=1}^N \langle J_i \rangle_i\ x_i^m \right)
\end{equation}
and the Bethe entropy
\begin{equation}
\label{eq:entropy}
S = \sum_{\mu=1}^M S_\mu - (N-1)\sum_{i=1}^N S_i\ 
\end{equation}
characterising the number of ``good'' coupling vectors. In the last
expression, the site entropy $S_i$ is given by
\begin{equation}
\label{eq:S_i}
S_i = - \sum_{J_i=-1,0,1} P_i(J_i)\ \ln P_i(J_i) \ ,
\end{equation}
and the pattern entropy
\begin{eqnarray}
\label{eq:S_mu}
S_\mu &=& - \sum_{\vec J} P_\mu(\vec J\; )\ \ln P_\mu(\vec J\; ) \\
P_\mu(\vec J\; ) &=& \exp\left\{ -\beta
\Theta\left[ -x_0^\mu\cdot \sum_k J_k x_k^\mu \right]\right\} 
\prod_{i}P_{i\to \mu}(J_i)
\nonumber
\end{eqnarray}
can be calculated in analogy to $\rho_{\mu\to i}$ via a Gaussian
approximation of the sum over $\vec J$.

So far, we did not fix the parameters $\beta$ and $h$ controlling the
energy and the dilution of the inferred vectors. A simple heuristic
strategy is the following: Fix the number $N_{eff}(\vec J\ )$
externally to some desired value, and start at high temperature and
initial $h$ given by $N_{eff} = 2 \cosh h / (1+2 \cosh h)$. Update the
messages starting from some random initialisation, and calculate the
energy of the average coupling vector and the Bethe entropy. Now
decrease the temperature gradually towards the zero-entropy (or zero
energy), adapting $h$ such that $N_{eff}(\vec J\ )$ remains close to
the desired value. This strategy can be repeated for various values of
$N_{eff}$. In practical applications, also an optimal $N_{eff}$ can be
determined by first dividing the data set into a training and a test
set, calculating the marginal probabilities on the training set, and
minimising the corresponding prediction error of the average coupling
vector on the test set. In the case of artificial data coming from a
known coupling vector, we can train on all data and compare directly
with the correct vector.

\section{Tests on artificial data}

Before coming to a theoretical analysis of the algorithmic performance,
we make some numerical test to get a first impression of the predictive 
power of the two algorithms.

\subsection{The data generator}

In order to check and compare the performance of the algorithms presented 
below, we have to introduce a data generator. It allows to create data sets with 
well-controlled properties, and the knowledge about the data-generating rule
allows to compare the results of the inference algorithm to the true rule.

In this work, we concentrate on data generated by a simple threshold function 
with couplings $J_i^0$,
\begin{equation}
\label{eq:generate}
x_0^\mu = {\rm sign} \left( \sum_{i=1}^N J_i^0 x_i^\mu + \eta^\mu \right)
\end{equation}
for all $\mu-1,...,M$. The inputs $\vec x^\mu = (x_1^\mu,...,x_N^\mu) $
have independently distributed binary entries.
With respect to the ternary classification into activators, repressors and 
non-interacting genes, this rule has two major differences
\begin{itemize}
\item {\it Heterogeneity:} The couplings $J_i$ may take values which are 
different from $0,\pm 1$. This allows to introduce regulatory variables of 
different interaction strengths, i.e. in particular weak and strong activators 
resp. repressors. In this work, we will study in particular the case where
$J_i \in \{0,\pm 1,\pm 2\}$. The couplings itself will be drawn randomly and 
independently with respect to a distribution
\begin{equation}
\rho(J_i^0) = (1-k_1-k_2) \delta(J_i^0) 
+ \frac{k_1}2 [ \delta(J_i^0+1) + \delta(J_i^0-1) ]
+ \frac{k_2}2 [ \delta(J_i^0+2) + \delta(J_i^0-2) ]
\end{equation}
Generalisations will be straight-forward, but they do not alter at all the
conclusions of this work. Note that real gene-regulatory networks are sparse, 
i.e. we have to work in 
the regime $k_{1,2}\ll 1$. In the statistical mechanics calculations presented 
below, we will consider the thermodynamic limit $N\to\infty$. At a first glance it
seems to be reasonable to assume the usual scaling of finite-connectivity
graphs, $k_{1,2} = {\cal O}(N^{-1})$,
for the coupling probabilities. Due to technical reasons we will, however, 
concentrate on the scaling  $k_{1,2} = {\cal O}(1) \ll 1$. This choice will be
justified by comparison with numerical data.
\item {\it Noise:} In biological data, there are various kinds of noise. The first 
one is biological noise resulting from the stochastic nature of cellular processes.
The second and more annoying one is experimental noise which is strongly present
in most high-throughput data. In addition, biological data are normally incomplete
in the sense, that not all existing gene expression levels are measured, 
that nutrient or mineral
concentrations, temperature and other external factors are not recorded etc.
In our data generator, we have included a simple additive noise term $\eta^\mu$ 
which we assume to be independent for various patterns $\mu$. More precisely we 
assume it to be Gaussian with
\begin{eqnarray}
\overline \eta^\mu &=& 0 \nonumber\\
\overline{ \eta^\mu \eta^\nu } &=& \gamma N \delta_{\mu,\nu} 
\end{eqnarray}
for all $\mu,\nu\in\{1,...,M\}$. Note that the scaling of the variance with 
$\sqrt{N}$  makes the noise to be of the same order of magnitude as the signal
$\sum J^0_i x_i^\mu$. For the special case $\gamma = k_1+4 k_2$, both noise 
and signal have exactly the same statistical properties.
\end{itemize}
Note that this data generation rule is more complex than the inference rule 
(\ref{eq:perz}). This will result in non-feasible data, i.e., a non-zero 
``energy'' ${\cal H}(\vec J)$ even for the best inferred couplings $\vec J$.
The best possible result would be $J_i = {\rm sign} J_i^0$ (here with 
${\rm sign(0)=0}$) since this would 
correctly assign the attributes activating, repressing, irrelevant to the
couplings. Due to the presence of noise as well as due to the finite amount 
of data available, such a perfect prediction will be impossible. We therefore
introduce the following notation
\begin{equation}
\begin{array}{lll}
J_i^0 = 0\ \ \ \  & J_i = 0\ \ \ \  & {\mbox{ true negative (TN)}}\\
J_i^0 \neq 0 & J_i = 0 & {\mbox{ false negative (FN)}}\\
J_i^0 = 0 & J_i \neq 0 & {\mbox{ false positive (FP)}}\\
J_i^0 \neq 0 & J_i \neq 0 & {\mbox{ true positive (TP)}}
\end{array}
\end{equation}
It can be refined taking also into account the relative sign of the
original and the inferred coupling, cf. the above discussion about
topology and functionality. One of the aims is to predict a {\it
fraction} of all couplings with high precision, i.e. to have an as
high as possible number of TP with a low number of FP. The quality
measure we use will be the confrontation of the {\it recall}, or {\it
sensitivity},
\begin{equation}
RC = \frac{N_{TP}}{N_{TP}+N_{FN}}
\end{equation}
and of the {\it precision}, or {\it specificity},
\begin{equation}
PR = \frac{N_{TP}}{N_{TP}+N_{FP}}\ .
\end{equation}
The recall describes the fraction of all existing non-zero couplings which are
predicted by the algorithm, whereas the precision tells us which fraction of
all predicted links is actually present in the data generator.

\subsubsection{Feasible data}
\label{sec:feasable_num}

As a first test, we have run BP on data sets with $k_2=\gamma=0$.
In this case, also the couplings of the data generator function assume
only the three values $J_i^0\in \{0,\pm 1\}$, and noise is absent. This means 
that the data set is feasible, the model (\ref{eq:perz}) on which we base our 
inference is able to reproduce the data without errors. We therefore work
automatically at zero formal temperature, and thus at zero energy.

The difficulty of the test problem comes from the fact that we are using an 
extremely sparse coupling vector: Only 3 out of $N=600$ input variables are 
coupled to the output via a non-zero coupling $J_i^0$. We also present 
relatively small data sets with $M\leq 70$. Is BP able to recognise the
three important input variables?

\begin{figure}[htb]
\vspace{0.2cm}
\begin{center}
\includegraphics[width=0.72\columnwidth]{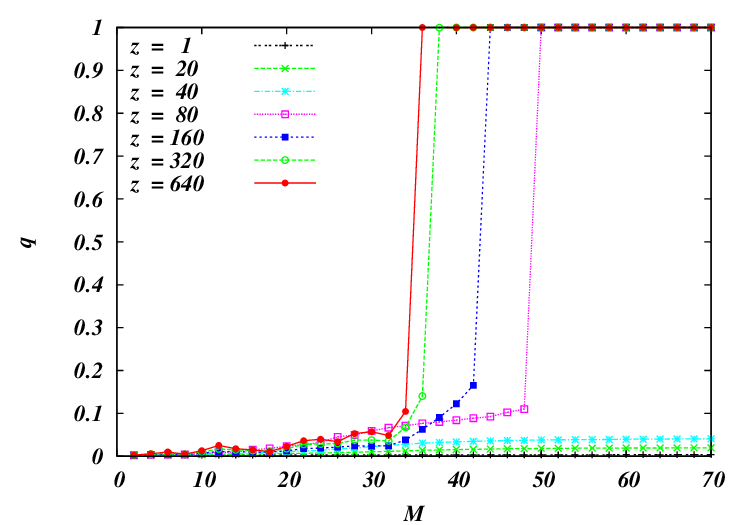}
\end{center}
\caption{Network inference from a feasible data set: 
Displayed are the overlap between the generator and the BP output as a function
of the number of presented patterns, for various values of $z=e^h$. The data 
generator has $N=600$ inputs, but only 3 are non-zero. Curves correspond to
one single test sample.}
\label{fig:martest}
\end{figure}

First, we have done experiments without the exterior diluting field $h$. The
results are given in Fig.~\ref{fig:martest}, where the quality of the prediction
is measured by the overlap
\begin{equation}
q = \frac 1{|\vec J^0|^2} \sum_{i=1}^N \sum_{J_i} J_i^0 J_i P_i(J_i)\ .
\end{equation}
This overlap takes for all considered numbers $M$ of presented patterns only 
very small values, the algorithm is not able to recognise the relevant
inputs. The situation changes drastically, if we add the diluting field
$h$. We find, at some $h$-dependent point, a discontinuous jump from a very bad
BP solution with small overlap to a perfectly polarised solution with $q=1$.
The algorithm has perfectly recognised the relevant three inputs, and thus
perfectly reconstructed the data generating rule! The number of needed patterns
decreases with $h$, and can be as small as $M=35$ for large fields. 

For even larger fields, the algorithm runs into problems: The dilution becomes 
as important for the algorithm as the energy constraints, and the BP 
solution polarises completely to the vector $J_i \equiv 0$. Best performance is 
obtained for fields just below this point.

\subsubsection{Unfeasible data}

\begin{figure}[htb]
\vspace{1cm}
\begin{center}
\includegraphics[width=0.64\columnwidth]{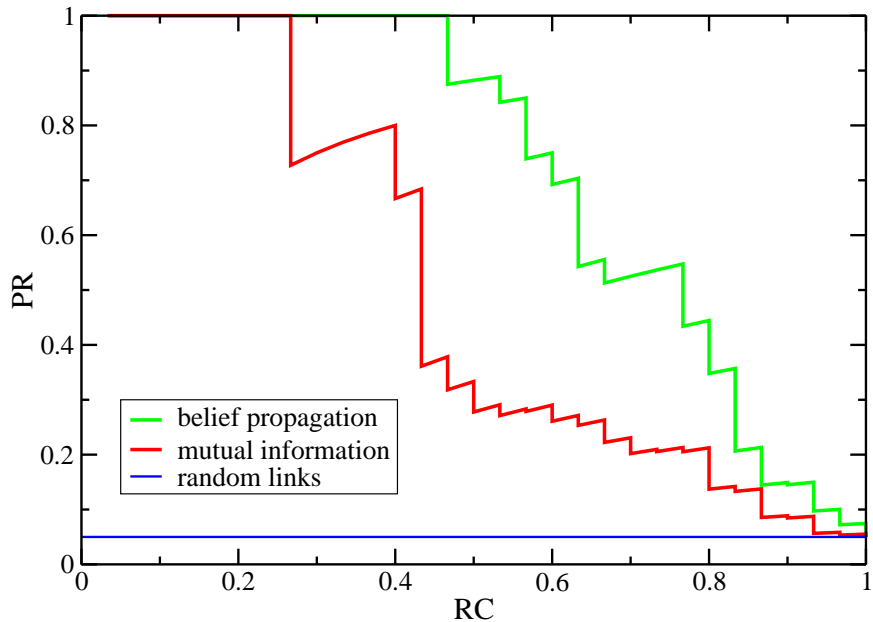}
\end{center}
\caption{Precision versus recall for inference with BP (green line) and
via pairwise mutual information (red line). 
The parameters of the model are $N=600, M=300$ and $k_1=k_2=0.025$, corresponding 
to 15 strong links ($J^0=\pm 2$) and 15 weak links ($J^0=\pm 1$), noise is
absent. The dilution value of BP is chosen to be slightly smaller than the
initial dilution, $n_{eff}=0.037$.}
\label{fig:performance}
\end{figure}

The success of BP for feasible data sets encourages us to go ahead to more
involved cases, i.e., cases with heterogeneities in the coupling strength
($k_1,k_2\neq 0$) and possibly with noise ($\gamma\neq 0$). Here we give a
first example of a data generator with $N=600$ inputs, but only 30 of them
have a non-zero coupling to the output - 15 of them with an absolute value
equal to two, 15 with absolute value one ($k_1=k_2=0.025$). We generate
$M=300$ patterns and apply both inference algorithms. The resulting rankings 
can be cut at arbitrary points, such that we can measure a recall-vs.-precision
curve as given in Fig.~\ref{fig:performance}.

We find that both algorithms start at precision one for the highest ranking
inputs. The algorithm based on mutual information includes, however, the
first false positive after 8 true positives, whereas BP is
able to find the first 14 true positives before including the first
erroneous link.  Also later on the performance of BP remains well above the
one of the MI. Note that most of the initially recognised links are strong
activators or repressors, only one of the first 14 true positives in BP is an
activator with $J_i^0=1$. Stronger signals are obviously detected first.

Even if not being perfect, it is to be noted that both algorithms perform
much better than  a random ranking which would fluctuate around an average
precision of only $PR=k_1+k_2=0.05$, as marked by the blue line in the figure.

\section{A statistical analysis of the algorithmic performance}
\label{sec:analysis}

Can we explain this behaviour analytically? Can we understand the
discontinuous transition from very bad to perfect inference in the
case of a feasible data set? Can we understand the relative behaviour
of BP compared to the different pair-correlation based methods?

In the following we will first calculate exactly the precision-versus-recall 
curve for the correlation coefficient and the mutual information, and than
we go to an analysis of the BP performance via a statistical mechanics analysis
of the Gibbs weight given in Eq.~(\ref{eq:j-distribution}). In both cases, we
are mainly interested in the thermodynamic limit $N\to\infty$. In this limit,
we also consider $M$ to diverge, with $\alpha=M/N$ being asymptotically
constant. The probabilities $k_1$ and $k_2$ for non-trivial couplings in
the data generator are considered to be finite, i.e. non-trivial couplings
constitute a finite fraction of all entries of $\vec J^0$.

\subsection{Pair correlation and mutual information}

Let us recall the estimate of the joint distribution of the values of
variable 0 and variable $i$ as calculated from the data. Using 
Eq.~(\ref{eq:joint}), and plugging in
Eq.~(\ref{eq:generate}) for the value of the outputs $x_0^\mu$, we get
\begin{equation}
\label{eq:joint2}
p(x_0,x_i) = \frac 1M \sum_{\mu=1}^M 
\delta(x_0,{\rm sign}[\vec J^0 \cdot\vec x^\mu+\eta^\mu])\ 
\delta(x_i,x_i^\mu)\ .
\end{equation}
Due to the random nature of the input patterns, this quantity itself is a
random variable with some distribution ${\mathbf P}(p(x_0,x_i))$.
Since the different patterns are statistically independent,
the joint probability results as a sum over $M=\alpha N$ independently
and identically distributed variables. According to the central limit theorem,
${\mathbf P}(p(x_0,x_1))$ is thus Gaussian. Its mean can be calculated as
\begin{eqnarray}
\overline{p(x_0,x_i)} &=& \frac 1{2^N} \sum_{x_1^1,...,x_N^1}
\delta(x_0,{\rm sign}[\vec J^0 \cdot\vec x^1+\eta^1])\ 
\delta(x_i,x_i^1)\nonumber\\
&=& \frac 1{2^N} \sum_{\{x_j^1;j\neq i\}}
\delta\left(x_0,{\rm sign}\left[J_i^0 x_i + \sum_{j\neq i} J_j^0 x_j^1
+\eta^1\right]\right)
\end{eqnarray}
where we have used that the input variables are independent and unbiased, 
and that the mean of all $M$ contributions to the sum is identical. For 
calculating this sum, we use another time the central limit theorem, the 
sum $y = \sum_{j\neq i} J_j^0 x_j^1$ is Gaussian distributed with mean zero
and second moment 
\begin{equation}
\overline{y^2} = (k_1+4 k_2+\gamma)N + {\cal O}(1) =: \Delta_y^2 \ .
\end{equation}
We thus get
\begin{eqnarray}
\overline{p(x_0,x_i)} &=& \frac 12 \int_{-\infty}^{\infty}
\frac {dy}{\sqrt{2\pi} \Delta_y} e^{-y^2/(2\Delta_y^2)}\ 
\delta\left(x_0,{\rm sign}[J_i^0 x_i + y]\right)
\nonumber\\
&=& \frac 12\ H\left( -\frac{J_i^0 x_0 x_i}{\sqrt{(k_1+4 k_2+\gamma)N}} 
\right)
\end{eqnarray}
with $H(x)=\int_x^\infty \frac{dy}{\sqrt{2\pi}} e^{-\frac{y^2}2}$. We 
expand this function around zero, and find
\begin{equation}
\label{eq:pmean}
\overline{p(x_0,x_i)} = \frac 14 + \frac{J_i^0 x_0 x_i}
{2\sqrt{2\pi (k_1+4 k_2+\gamma)N}} + {\cal O}(N^{-1})\ .
\end{equation}
The calculation of the variance is now straight-forward. We have
\begin{equation}
\overline{ p(x_0,x_i)^2 } = \frac{M-1}{M}\ \overline{ p(x_0,x_i) }^2
+ \frac 1M \ \overline{ p(x_0,x_i) }\ ,
\end{equation}
and consequently we find
\begin{eqnarray}
\label{eq:pvar}
\overline{ [\Delta p(x_0,x_i)]^2 } &=& \frac{1}{M}\ \overline{ p(x_0,x_i) }
\ (1- \overline{ p(x_0,x_i) }) \nonumber\\
&=& \frac 3{16\alpha N} + {\cal O}( N^{-3/2} )\ .
\end{eqnarray}
Finally we need the correlation for different value pairs
$(x_0,x_i)\neq(x'_0,x'_i)$. Following the the same strategy as before,
we find
\begin{equation}
\label{eq:pcorr}
\overline{ \Delta p(x_0,x_i)\ \Delta p(x'_0,x'_i) } 
= - \frac 1{16\alpha N} + {\cal O}( N^{-3/2} )\ .
\end{equation}
A side remark is necessary here. The covariance matrix of all four
values of $p(x_0,x_i)$ is degenerate, with a zero-eigenvector $(1,1,1,1)$.
It is, however, clear that all four quantities cannot have a simple joint
Gaussian distribution - given three, the fourth is determined by
normalisation, $\sum_{x_0,x_i} p(x_0,x_i) = 1$. The zero-eigenvector thus 
corresponds to the forbidden normalisation breaking fluctuations.

\subsubsection{Pair correlations}

Let us, however continue with the pair correlations between the output 
variable $x_0$ and a potential regulator $x_i$. We write
\begin{eqnarray}
C_{0i} &=& \sum_{x_0,x_i} p(x_0,x_i)\ x_0 x_i \nonumber\\
&=& p(1,1)+p(-1,-1)-p(1,-1)-p(-1,1) \nonumber\\
&=& 2\ p(1,1)+2\ p(-1,-1) -1
\end{eqnarray}
where, in the last line, we have exploited the normalisation of $p(x_0,x_1)$. 
Due to imperfect sampling in a finite data set, this is again a Gaussian random 
variable. Using Eq.~(\ref{eq:pmean}) we find its mean to be
\begin{equation}
\label{eq:cmean}
\overline{C_{0i}} = \frac{2 J_i^0} {\sqrt{2\pi (k_1+4 k_2+\gamma)N}} 
+ {\cal O}(N^{-1})\ ,
\end{equation}
whereas its variance can be calculated directly from Eqs.~(\ref{eq:pvar}) and 
(\ref{eq:pcorr}),
\begin{equation}
\label{eq:cvar}
\overline{ (\Delta C_{0i})^2 } 
= \frac 1{\alpha N} + {\cal O}( N^{-3/2} )\ .
\end{equation}
Both mean and variance scale as ${\cal O}(N^{-1/2})$, introducing thus 
$\tilde C_{0i}=\sqrt N\ C_{0i}$ we find its probability distribution
\begin{equation}
\label{eq:cdistr}
P(\tilde C_{0i} | J_i^0) = \sqrt{\frac \alpha{2\pi}} \exp\left\{
-\frac \alpha 2 \left( \tilde C_{0i} - \frac{2 J_i^0}
{\sqrt{2\pi (k_1+4 k_2+\gamma)}} 
\right)^2 \right\}
\end{equation}
conditioned to the value of the coupling $J_i^0$ between variables $0$ and 
$i$ in the data generator (\ref{eq:generate}). At this point, one recognises 
the potential and also the limitations of network inference via pair
correlations: The mean value of the correlations of different variables is 
well separated by a term being proportional to the coupling strength, and
in particular strong activators and inhibitors should be distinguishable
from irrelevant variables. On the other hand, the variance of distribution
(\ref{eq:cdistr}) is of the same order of magnitude as the mean, i.e. the
distributions for different values $J_i^0$ overlap, and errors in the 
network inference are unavoidable. The only way to suppress errors is to
augment the amount of data, the variance of the measured correlation values
around its expectation decays as $1/\sqrt\alpha$.

The measured pair correlations, or more precisely their absolute values,
serve as a ranking for the inclusion of different variables into the
reconstructed network. Let us therefore imagine that we introduce 
a cutoff $\theta>0$ for the absolute 
value of the rescaled correlations $C_{0i}$, i.e., for all $i$ with 
$|\tilde C_{0i}|>\theta$ we introduce a coupling, for all inputs with a 
smaller absolute correlation value we assume the coupling to be absent.
The fraction of included couplings with given original $J^0$ is than found
to be
\begin{eqnarray}
P\left(\ |\tilde C_{0i}|>\theta\ |\ J^0 \right) &=&
\int_{-\infty}^{-\theta} d\tilde C_{0i} P(\tilde C_{0i} | J^0)
+ \int_{\theta}^{\infty} d\tilde C_{0i} P(\tilde C_{0i} | J^0)
\nonumber\\
&=& H\left( \sqrt\alpha \left[ \theta - \frac{2 J_i^0}
{\sqrt{2\pi (k_1+4 k_2+\gamma)}} \right]\right) +
 H\left( \sqrt\alpha \left[ \theta + \frac{2 J_i^0}
{\sqrt{2\pi (k_1+4 k_2+\gamma)}} \right]\right)
\end{eqnarray}
For $J^0\neq 0$, the corresponding included couplings are true positives.
For $J^0= 0$, they are false positives. We thus can write up the recall and
the precision of the inferred network, in dependence of the cutoff parameter
$\theta$:
\begin{eqnarray}
\label{eq:recall-corr}
RC(\theta) &=& 
\frac{k_1 P\left(\ |\tilde C_{0i}|>\theta\ |\ \pm 1 \right)
+ k_2 P\left(\ |\tilde C_{0i}|>\theta\ |\ \pm 2 \right) }{k_1+k_2}
\nonumber\\
PR(\theta) &=& 
\frac{k_1 P\left(\ |\tilde C_{0i}|>\theta\ |\ \pm 1 \right)
+ k_2 P\left(\ |\tilde C_{0i}|>\theta\ |\ \pm 2 \right) }
{(1-k_1-k_2) P\left(\ |\tilde C_{0i}|>\theta\ |\ 0 \right)
+ k_1 P\left(\ |\tilde C_{0i}|>\theta\ |\ \pm 1 \right)
+ k_2 P\left(\ |\tilde C_{0i}|>\theta\ |\ \pm 2 \right) }
\end{eqnarray}
Changing the value of $\theta$ from high to small values, we increase the
recall from zero to one, at the cost of a decreasing precision. More
details will be given below, in the comparison to the results of the
massage-passing based inference algorithm.

Note that we did not require the correct determination of the sign of
the coupling, a generalisation to this case is, however, straight forward.
For not too small $\alpha$, it practically does not lead to recognisable
differences, i.e. the waste majority all true positives is recognised due
to the correct sign of the measured correlation.

\subsubsection{Mutual information}

In how far does the network reconstruction change if we go from
correlations to the pairwise mutual information as given in Eq.~(\ref{eq:mi})? 
The latter has a nonlinear dependence on the pair correlations $p(x_0,x_1)$,
and consequently it does not have a Gaussian distribution. We may, however,
use the fact that variables are only weakly correlated, and write
\begin{equation}
\label{eq:zcorr}
p(x_0,x_i) = \frac 14 + \frac {z(x_0,x_i)}{\sqrt N}\ .
\end{equation} 
The newly introduced function $z(x_0,x_1)$ contains all non-trivial
correlations between $x_0$ and $x_1$, and it is of ${\cal O}(N^0)$. 
Note that at this point we consider Eq.~(\ref{eq:zcorr}) to be an
exact definition of $z(x_0,x_i)$, i.e. the latter contains all the
lower-order corrections in $N$ and not only the dominant term. Due to the
normalisation of $p$ it fulfils $\sum_{x_0,x_i} z(x_0,x_i)=0$. We further
on write for the two marginal single-variable distributions of 
$p(x_0,x_i)$:
\begin{eqnarray}
p(x_0) &=& \frac 12 + \frac {z(x_0)}{\sqrt N} \nonumber\\
p(x_i) &=& \frac 12 + \frac {z(x_i)}{\sqrt N}\ .
\end{eqnarray}
Plugging this into the definition of the mutual information, we find
\begin{equation}
I_{0i} = \sum_{x_0,x_i} \left[\frac 14 + \frac {z(x_0,x_i)}{\sqrt N}
\right] \ \left[ \log\left( 1+ \frac {4\ z(x_0,x_i)}{\sqrt N}\right)
- \log\left( 1+ \frac {2\ z(x_0)}{\sqrt N}\right)
- \log\left( 1+ \frac {2\ z(x_i)}{\sqrt N}\right)
\right]
\end{equation}
Expanding the logarithm up to second order in the $z$, and using the
normalisation to eliminate all linear terms, we find after some 
straight-forward algebra
\begin{eqnarray}
I_{0i} &=& \frac 2N \left[ z(1,1)+z(-1,-1)
\right]^2 + {\cal O}\left( N^{-\frac 32} \right) \nonumber\\
&=& \frac {C_{01}^2}2 + {\cal O}\left( N^{-\frac 32} \right) \ .
\end{eqnarray}
We thus conclude that, at least under the scaling limits considered in this
work, the correlation coefficient and the mutual information carry the same
information. A network reconstruction algorithm based on either of these two
quantities thus leads to the same results.

\subsection{Characterising the set of coupling vectors}

In the last subsection, we have analytically characterised the performance of
a network inference algorithm based on pairwise correlation measures. Is a
similar comparison possible for the belief-propagation algorithm introduced in
Sec.~\ref{sec:bp}? As already mentioned, the algorithm is based on the measure
\begin{equation}
P(\vec J) = \frac 1{Z(\beta,h)} 
\exp\left\{ -\beta {\cal H}(\vec J) - h N_{eff}(\vec J) \right\}
\end{equation}
which assigns a weight to every coupling vector $\vec J$. This weight depends 
on its performance on the data set $\{\vec x^\mu\}$ as measured by the 
Hamiltonian ${\cal H}(\vec J)$ defined in Eq.~(\ref{eq:hamiltonian}), and on the
effective number $N_{eff}(\vec J)$ of relevant inputs $J_i\neq 0$ as defined in 
Eq.~(\ref{eq:neff}). A positive inverse temperature $\beta$ introduces a bias 
toward $\vec J$ with few errors on the data set, a positive $h$ favours diluted
coupling vectors. The aim of the algorithm was
to determine the marginal distributions $P_i(J_i)$ for single entries of the
vector $\vec J$, and use the probability $P_i(J_i\neq 0)$ that such a component
becomes non-zero for ranking the input variables according to their relevance
for the output. 

Note that in this way we do not work with a single reconstructed network, which 
may or may not be similar to the data generator, but with a probability measure
over all feasible coupling vectors.

To characterise the performance of this ranking procedure we have to calculate
first the partition function
\begin{equation}
Z(\beta,h) = \sum_{\vec J\in \{0,\pm1\}^N} 
\exp\left\{ -\beta {\cal H}(\vec J) - h N_{eff}(\vec J) \right\}\ .
\end{equation}
From this we can calculate the average energy
\begin{equation}
\langle {\cal H} \rangle = \varepsilon N = - \frac \partial{\partial \beta}
\log Z(\beta,h)
\end{equation}
and the average number of non-zero couplings
\begin{equation}
\langle N_{eff} \rangle = n_{eff} N = - \frac \partial{\partial h}
\log Z(\beta,h)\ .
\end{equation}
The entropy characterising the number of statistically dominant coupling vectors
is now given by
\begin{equation}
S = s N = \log(Z) + \beta \langle {\cal H}\rangle + h \langle N_{eff} \rangle\ .
\end{equation}
Besides these standard thermodynamic quantities, we also need the
distributions $P(J\neq 0\ | \ J^0)$, i.e., the probability that a
single coupling component is different from zero, given the
corresponding coupling in the data generator.  This distribution will
allow us to identify true and false positives following a similar
strategy as in the last section. The calculation necessary to obtain
these thermodynamic quantities is a modified Gardner calculation as
introduced in the theory of formal neuronal networks. This calculation
follows a path beautifully reviewed in the text book by Engel and van
den Broeck \cite{EnBr}, we therefore skip all technical details and
present only the major steps.

A first observation is that the free energy $-\beta^{-1}\log Z(\beta,
h)$ depends explicitly on the realisation of the input data, i.e. on
the $\vec x^\mu$, the noise $\eta^\mu$ and the coupling vector $\vec
J^0$ of the data generator. It is consequently a random variable which
is, however, expected to be self-averaging: In the thermodynamic
limit, its distribution is expected to become more and more
concentrated around its average, and relative fluctuations are
expected to disappear. We can therefore replace the actual free energy
by its mean value $-\beta^{-1}\overline{\log Z(\beta, h)}$, where the
overline denotes the joint average over $\vec x^\mu$,
$\eta^\mu$ and $\vec J^0$. Technically this can be achieved using the replica trick:
We write the logarithm as
\begin{equation}
\overline{\log Z} = \lim_{n\to 0} \frac{\overline{Z^n}-1}n\ ,
\end{equation}
calculate the right-hand side first for positive integer $n$, introduce an
analytical continuation to real $n$ and perform the limit. The average over 
$Z^n$ for integer $n$ can be performed easily since the $n$-th power can be
interpreted as an $n$-fold identical replication of the original system.

We thus have to calculate the following:
\begin{eqnarray}
\overline{Z^n} &=&  2^{-MN} \sum_{x_i^\mu} \int \prod_{\mu=1}^N
\left[ d \eta^\mu e^{- (\eta^\mu)^2/(2\gamma N)} \right] 
\sum_{\{\vec J^a; a=0,...,n\}} \prod_{i=1}^N \left[ \rho(J_i^0)
e^{-h \sum_{a=1}^n |J_i^a|} \right] \nonumber\\
&&\times \exp\left\{ -\beta \sum_{a=1}^n \sum_{\mu=1}^M 
\Theta\left( - \left[\sum_{i=1}^N J_i^0 x_i^\mu + \eta^\mu\right]
\left[\sum_{i=1}^N J_i^a x_i^\mu \right]\ .
\right)\right\}
\end{eqnarray}
Here, $a=1,...,n$ enumerates the $n$ replicas of the system, whereas
$a=0$ stands for the data generator. 
Following the path explained in \cite{EnBr}, we can extract the
interesting quantities under the assumption of replica-symmetry:
\begin{eqnarray}
\label{eq:rsresult}
\varepsilon &=& 2\alpha \int Dy\ 
H\left[ y \sqrt{\frac{r^2}{k^0 q-r^2}} \right]\ 
\frac {e^{-\beta} H\left[-y \sqrt{\frac{q}{n_{eff}-q}}\right]}
{1+(e^{-\beta}-1) H\left[-y \sqrt{\frac{q}{n_{eff}-q}}\right]}\nonumber\\
s &=& -n_{eff}\hat k -r \hat r + \frac 12 q\hat q 
+\int Dx\ \sum_{J^0} \rho(J^0) \log\left[ 1+2e^{\hat k+ \hat q/2}
\cosh\{ \hat r J^0 +x \sqrt{\hat q} \}\right] \\
&& +2\alpha\int Dy\ H\left[ y \sqrt{\frac{r^2}{k^0 q-r^2}} \right]\ 
\left\{ \log\left(1+(e^{-\beta}-1) H\left[-y \sqrt{\frac{q}{n_{eff}-q}}
\right]\right)
+\beta \frac {e^{-\beta} H\left[-y \sqrt{\frac{q}{n_{eff}-q}}\right]}
{1+(e^{-\beta}-1) H\left[-y \sqrt{\frac{q}{n_{eff}-q}}\right]}
\right\} \nonumber
\end{eqnarray}
with the Gaussian normal distribution $Dy=dy/\sqrt{2\pi}\ e^{-y^2/2}$,
$k^0=k_1+4k_2+\gamma$
and the order parameters being determined via the self-consistent
saddle-point equations
\begin{eqnarray}
n_{eff} &=& \frac 1N \langle \vec J^a\cdot\vec J^a \rangle\nonumber\\
&=& \int Dx\ \sum_{J^0} \rho(J^0) \frac
{2e^{\hat k+ \hat q/2} \cosh\{ \hat r J^0 +x \sqrt{\hat q} \}}
{1+2e^{\hat k+ \hat q/2} \cosh\{ \hat r J^0 +x \sqrt{\hat q} \}}
\nonumber\\
q &=& \frac 1N \langle \vec J^a\cdot\vec J^b \rangle\nonumber\\
&=& \int Dx\ \sum_{J^0} \rho(J^0) \left[\frac
{2e^{\hat k+ \hat q/2} \sinh\{ \hat r J^0 +x \sqrt{\hat q} \}}
{1+2e^{\hat k+ \hat q/2} \cosh\{ \hat r J^0 +x \sqrt{\hat q} \}}
\right]^2 \nonumber\\
r &=& \frac 1N \langle \vec J^0\cdot\vec J^a \rangle\nonumber\\
&=& \int Dx\ \sum_{J^0} \rho(J^0)\, J^0\, \frac
{2e^{\hat k+ \hat q/2} \sinh\{ \hat r J^0 +x \sqrt{\hat q} \}}
{1+2e^{\hat k+ \hat q/2} \cosh\{ \hat r J^0 +x \sqrt{\hat q} \}} \\
\hat q &=& -\frac {\alpha k^0r}{\pi(k^0q-r^2)^{3/2}}
\int dy\ y\ exp\left\{ - \frac {k^0 q y^2}{2(k^0q-r^2)}\right\} 
\log\left(1+(e^{-\beta}-1) H\left[-y \sqrt{\frac{q}{n_{eff}-q}}
\right]\right) \nonumber\\
&&+\frac {\alpha k}{\pi\sqrt q (n_{eff}-q)^{3/2}}
\int dy\ y\ exp\left\{ - \frac {k y^2}{2(n_{eff}-q)}\right\} 
\frac {e^{-\beta} H\left[-y \sqrt{\frac{q}{n_{eff}-q}}\right]}
{1+(e^{-\beta}-1) H\left[-y \sqrt{\frac{q}{n_{eff}-q}}\right]} \nonumber\\
\hat r &=& -\frac {\alpha k^0r}{\pi(k^0q-r^2)^{3/2}}
\int dy\ y\ exp\left\{ - \frac {k^0 q y^2}{2(k^0q-r^2)}\right\} 
\log\left(1+(e^{-\beta}-1) H\left[-y \sqrt{\frac{q}{n_{eff}-q}}
\right]\right) \nonumber
\end{eqnarray}
The parameter $n_{eff}$ measures the self-overlap of a typical
coupling vector with itself, and thus the desired fraction of non-zero
components, $q$ measures the overlap between two different coupling
vectors, $r$ the overlap between a typical coupling vector and the
coupling of the original data generating vector $\vec J^0$, and their
conjugate parameters. All these parameters still depend on $\beta$ and 
$h$. As explained in the algorithmic section, we fix them by requiring
$n_{eff}$ to be in some small given interval, and by increasing the temperature
such that either the energy or the entropy go to zero, focusing in this
way to the ground states at given $n_{eff}$.

At this point, we are also able to express the probability 
$P(J\neq 0|J^0)$ that a link is present in an inferred coupling vector,
in dependence on its value in the data generator. We find that it takes
the value
\begin{equation}
\label{eq:pjj}
P(J\neq 0 | J^0) = \frac
{2e^{\hat k+ \hat q/2} \cosh\{ \hat r J^0 +x \sqrt{\hat q} \}}
{1+2e^{\hat k+ \hat q/2} \cosh\{ \hat r J^0 +x \sqrt{\hat q} \}}
\end{equation}
with $x$ being a random variable drawn from a Gaussian normal
distribution. This random number represents the heterogeneity of the
different inputs having the same value of the coupling in the data
generator, which persists due to the imperfect sampling at finite
$\alpha$. Accepting all those inputs having a value of $P(J\neq
0|J^0)$ larger than an arbitrary threshold $\phi\in(0,1)$, we find the
fraction of  accepted couplings to equal
\begin{eqnarray}
{\mathbf P}\left[ P(J\neq 0 | J^0)>\phi \right]
&=&\;\;\, H\left[ \frac 1{\sqrt{\hat q}} \left(-\sqrt{\hat r}J^0+ {\rm acosh} 
\frac{2e^{\hat k+ \hat q/2}(1-\phi)}\phi
\right)\right] \nonumber\\&&+
H\left[ \frac 1{\sqrt{\hat q}} \left(+\sqrt{\hat r}J^0+ {\rm acosh} 
\frac{2e^{\hat k+ \hat q/2}(1-\phi)}\phi
\right)\right]
\end{eqnarray}
In strict analogy to Eqs.~(\ref{eq:recall-corr}), we can determine the
recall and the precision of the inferred network, in dependence of the
cutoff parameter $\phi$:
\begin{eqnarray}
RC(\phi) &=& 
\frac{k_1 {\mathbf P}\left[ P(J\neq 0 | \pm 1)>\phi \right]
+ k_2 {\mathbf P}\left[ P(J\neq 0 | \pm 2)>\phi \right] }{k_1+k_2}
\nonumber\\
PR(\phi) &=& 
\frac{k_1 {\mathbf P}\left[ P(J\neq 0 | \pm 1)>\phi \right]
+ k_2 {\mathbf P}\left[ P(J\neq 0 | \pm 2)>\phi \right] }
{(1-k_1-k_2) {\mathbf P}\left[ P(J\neq 0 | 0)>\phi \right]
+ k_1 {\mathbf P}\left[ P(J\neq 0 | \pm 1)>\phi \right]
+ k_2 {\mathbf P}\left[ P(J\neq 0 | \pm 2)>\phi \right] }
\end{eqnarray}
Again, for a very strict acceptance criterion $\phi$ slightly below
one, we start with a small recall, but high precision. Approaching
zero with the cutoff, the recall becomes better and better, but the
precision goes down.

\subsection{The algorithmic performance}

\subsubsection{Feasible data and perfect reconstruction}

\begin{figure}[htb]
\vspace{0.5cm}
\begin{center}
\includegraphics[width=0.65\columnwidth]{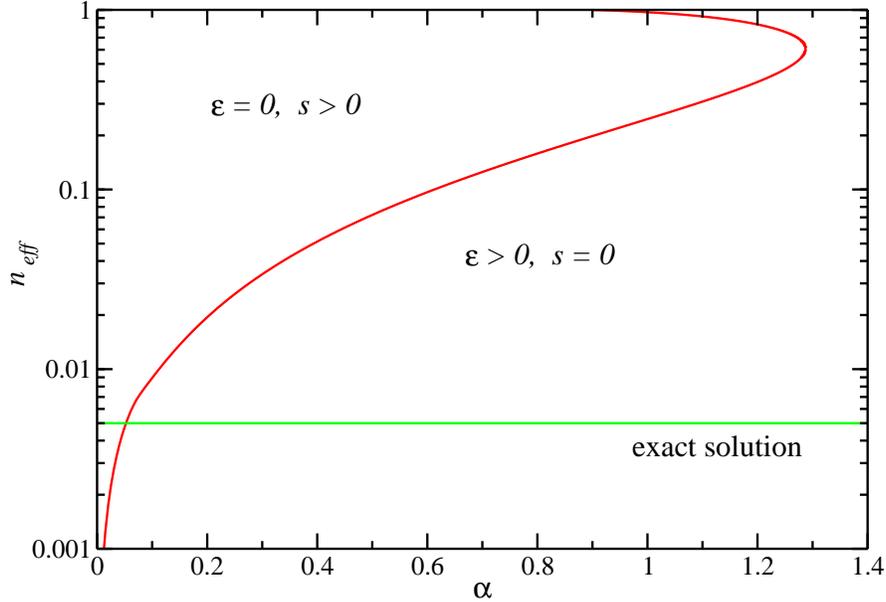}
\end{center}
\caption{Phase diagram for a feasible data set with $k_1=0.005,\
k_2=0,\ \gamma=0$. The red line separates the phase space region with
error-less solutions from a region where the ground-state energy is
positive. The always existing perfect solution $\vec J=\vec J^0$ is
given by the green line.}
\label{fig:phasediagram_K005}
\end{figure}

In Sec.~\ref{sec:feasable_num} we have seen that, in the case of a
feasible data set, a discontinuous jump from a region of bad to
perfect network inference appears. This jump can be easily understood
on the basis of the phase diagram displayed in
Fig.~\ref{fig:phasediagram_K005}. If we have a relatively small number
of input patterns ($\alpha<1.287$ in the parameter choice of the
figure), there is a large variety of $\vec J$ vectors perfectly
reproducing the data set. The point is, however, that almost all of
them are very dense, i.e., they have a large $n_{eff}$. Without an 
external field favouring diluted couplings, the algorithm automatically
concentrates to the maximum entropy point at high $n_{eff}$ (which we
found numerically to be always at $n_{eff}\geq 0.5$), and the 
quality of the reconstructed network is very bad. Increasing the diluting
external field, the value of $n_{eff}$ decreases first continuously,
remaining still in the region of positive entropy and zero energy.  At
a certain field we hit, however, the red line in
Fig.~\ref{fig:phasediagram_K005}. It marks the phase transition to a
region where ground states have positive energy and zero entropy. Our
algorithm is, however, forced to work at zero energy, it jumps
therefore to the next diluted solution with $\varepsilon=0$ - which is
the perfect solution $\vec J=\vec J^0$. The algorithm is thus able to
perfectly infer the data generator as soon as we have a value of
$\alpha=M/N$ being larger than the one of the intersection point of
the phase boundary and the perfect solution line (More precisely, this
point is a lower bound since BP stays a while at a metastable solution 
with negative entropy before jumping to the perfect solution). Using 
the parameters of the example in Sec.~\ref{sec:feasable_num} and 
Fig.~\ref{fig:phasediagram_K005}, we find that for perfect reconstruction 
we need to have $\alpha=M/N\geq 0.053$, for $N=600$ potential inputs
this corresponds to at least 32 patterns. Note that this value is
consistent with the single-sample experiment in Sec.~\ref{sec:feasable_num}.

The existence of a first order transition underlines the necessity to 
introduce the diluting field $h$. Without the field, one would have to work 
at $\alpha>1.287$, which, in the case shown in Sec.~\ref{sec:feasable_num}, 
would correspond to more than 770 patterns. With the use of the field, we
have perfect generalisation already with slightly more than 32 patterns. 
This increase in predictive power is impressive, remembering in particular
that the diluting field was biologically motivated by the fact that 
gene-regulatory networks are sparse.

\subsubsection{Unfeasible data and partial reconstruction}

As already discussed before, the feasible situation is very
unrealistic, and therefore only a first (but important) test case. In
the following, we are going to investigate the behaviour for
heterogeneous and possibly noisy data generators.

In this section, we therefore consider data coming from a generator with
couplings $J_i^0\in\{0,\pm1,\pm2\}$, but try to infer them on the basis
of our ternary classification to activators, repressors and irrelevant
inputs. The more complex nature of the data generator leads to the fact
that no perfect zero-energy solution exists any more, and we have to go
to finite energies (resp. finite temperatures) in the algorithm. As 
explained before, the parameters are fixed such that we work at various
preselected dilutions, and the temperature is chosen such that we are in
a corresponding ground state, {\it i.e.}, we work at zero entropy.

More precisely, in all the rest of the chapter we concentrate on the
parameter selection $k_1=k_2=0.025$, {\it i.e.}, 2.5\% of all relevant
links are weak activators or repressors, and other 2.5\% are strong
couplings. The remaining 95\% are irrelevant inputs.

\begin{figure}[htb]
\vspace{1.5cm}
\begin{center}
	\includegraphics[width=0.9\columnwidth]{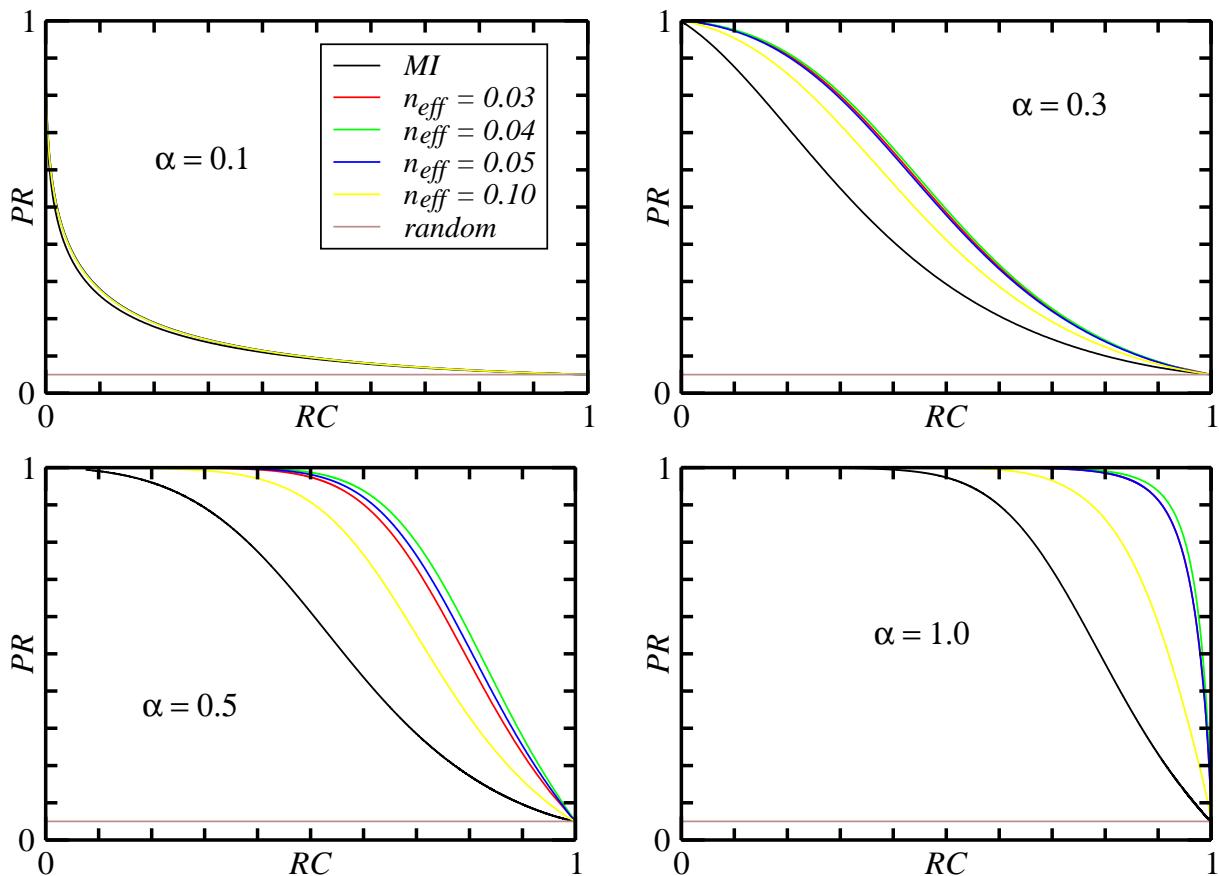}
\end{center}
\vspace{1cm}
\caption{Theoretical results of the performance of BP versus
pair correlation based methods like ARACNE, for various values of 
$\alpha=M/N$.}
\label{fig:bp_vs_aracne}
\end{figure}

\begin{figure}[htb]
\vspace{1.5cm}
\begin{center}
	\includegraphics[width=0.6\columnwidth]{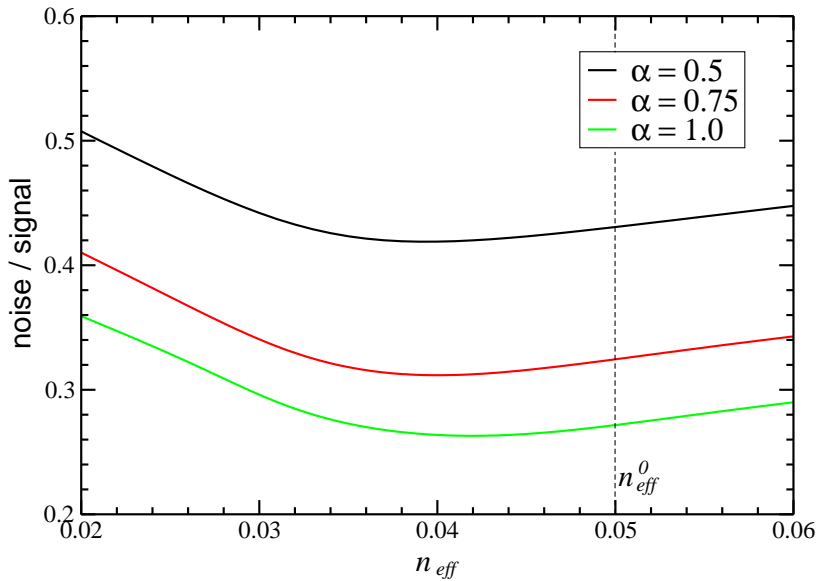}
\end{center}
\vspace{0.2cm}
\caption{Inverse signal-to-noise ratio: Best inference at given
data-set size $\alpha$ is possible at the minimum of the curve.}
\label{fig:peak_separ}
\end{figure}

Let us first concentrate a moment to the error-less case, $\gamma=0$,
and investigate the influence of the number $\alpha$ of presented
patterns per input bit. In Fig.~\ref{fig:bp_vs_aracne} we have given
the results for the precision-vs.-recall curves at various dilutions.
If the density of the inferred network is set too high ($n_{eff}=0.1$
in the figure), the performance is not very good, but it increases if
we further dilute the inferred network. The best performance is
obtained if the dilution of the inferred network is slightly stronger
than the one of the data generator. The reason for this finding is
very clear: In this way the algorithm keeps the strongest good
signals, but cuts many of the strongest false positive signals -
obviously at the cost of introducing some false negatives. A stronger
diluting field would cut out more true positives becoming thereby
false negatives, a weaker field would include more false positives
which before were true negatives. A more formal reason can be seen
looking to Eq.~(\ref{eq:pjj}): In this equation the effective signal
due to a coupling $J^0$ is given by $\hat r J^0$, and it is distorted
by the noise $x \sqrt{\hat q}$, with $x$ being normally distributed.
The best inference is possible if diverse signals are maximally
separable, {\it i.e.}, if the noise-to-signal ratio $\sqrt{\hat
q}/\hat r$ is minimal. In Fig.~\ref{fig:peak_separ}, we plot this
ratio for various $\alpha$, as a function of the effective dilution
$n_{eff}$. The optimal value is found at $n_{eff}<n^0_{eff}$: We also
see that the position of the minimum moves toward $n_{eff}^0$ for
growing $\alpha$.

In Fig.~\ref{fig:bp_vs_aracne}, we have also included the
corresponding curves for algorithms like ARACNe based on pairwise
correlations - the message-passing approach takes into account
collective effects and therefore clearly outperforms the local
pair-correlation methods.

\begin{figure}[htb]
\vspace{0.5cm}
\begin{center}
\includegraphics[width=0.65\columnwidth]{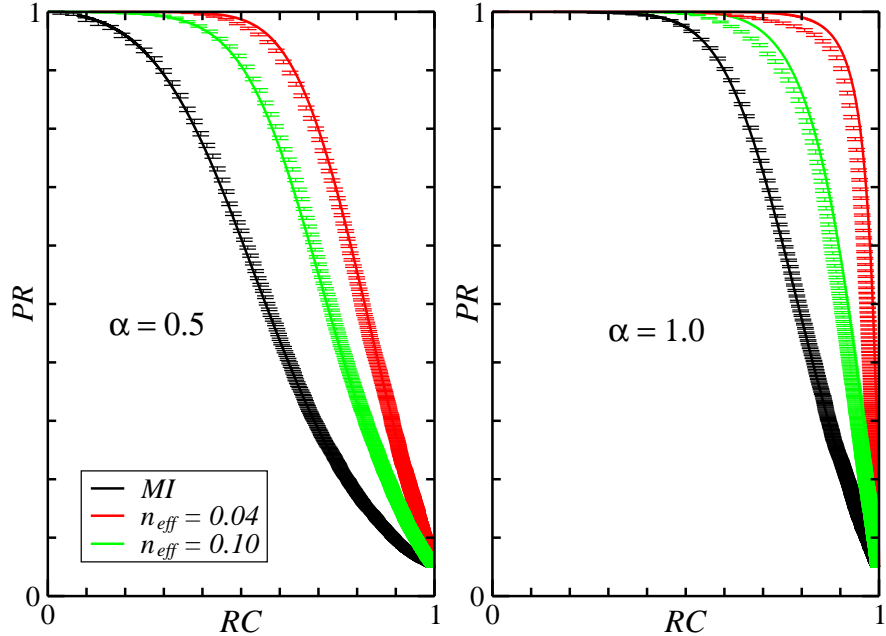}
\end{center}
\vspace{0.2cm}
\caption{Theoretical versus numerical results. The numerical results
are produced at $N=600$, and averaged over 500 realisations of the 
disorder.}
\label{fig:theory_vs_numerics}
\end{figure}

In Fig.~\ref{fig:theory_vs_numerics} we have compared the analytical 
findings with the performance of BP on single samples of size $N=600$,
averaged over 500 samples. Even if the algorithm includes a Gaussian
approximation compared to the exact BP equations, the coincidence of
both approaches is very good.

\begin{figure}[htb]
\vspace{1.5cm}
\begin{center}
\includegraphics[width=0.65\columnwidth]{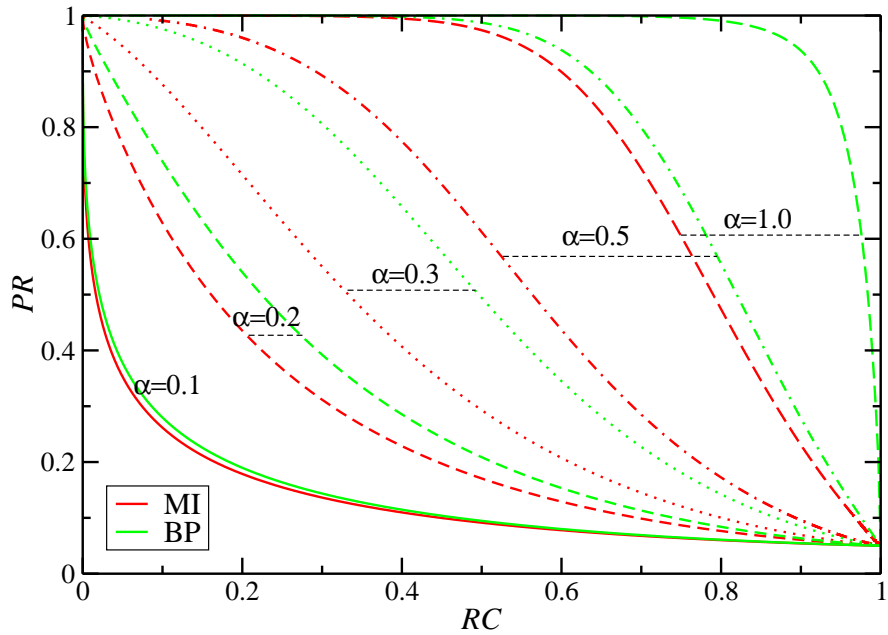}
\end{center}
\caption{Increase in performance with $\alpha$, i.e. with the number
of presented patterns.}
\label{fig:precision_alpha}
\end{figure}

\begin{figure}[htb]
\vspace{1.5cm}
\begin{center}
	\includegraphics[width=0.65\columnwidth]{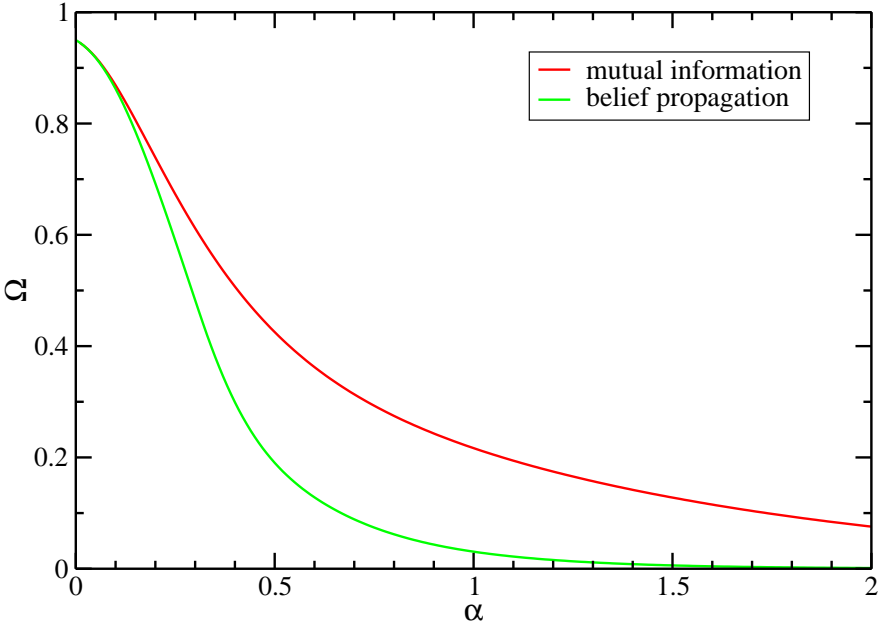}
\end{center}
\caption{Prediction error as a function of the number $M=\alpha N$
of presented patterns, for both algorithms based on pairwise mutual
information (red line) and belief propagation (green line, measured 
at constant dilution $n_{eff}=0.04$.}
\label{fig:prediction_error}
\end{figure}

Another striking finding of Fig.~\ref{fig:bp_vs_aracne} is the strong
improvement of the predictive quality with increasing data sets: For
$\alpha=0.1$, the precision starts to decrease immediately with 
increasing recall, whereas it stays practically equal to one for
higher values of $\alpha$, cf. also Fig.~\ref{fig:precision_alpha}. 
This observation can be quantified better by measuring the prediction 
error
\begin{equation}
\Omega = 1-\int_0^1 PR \ dRC 
\end{equation}
which gives the area over the precision-vs.-recall curve. The
prediction error is zero for a perfect predictor, and goes to
$1-k_1-k_2=0.95$ for a completely random predictor. In
Fig.~\ref{fig:prediction_error} we have plotted the prediction error
as a function of the size of the data set as given by $\alpha$: The
curves for both algorithm types start obviously at $\Omega=0.95$ for
empty data sets ($\alpha=0$), without data there cannot be any
prediction. The error than decreases exponentially fast with growing
$\alpha$, but we see that BP behaves much better than the curves
corresponding to the mutual information.

\begin{figure}[htb]
\vspace{1.5cm}
\begin{center}
\includegraphics[width=0.65\columnwidth]{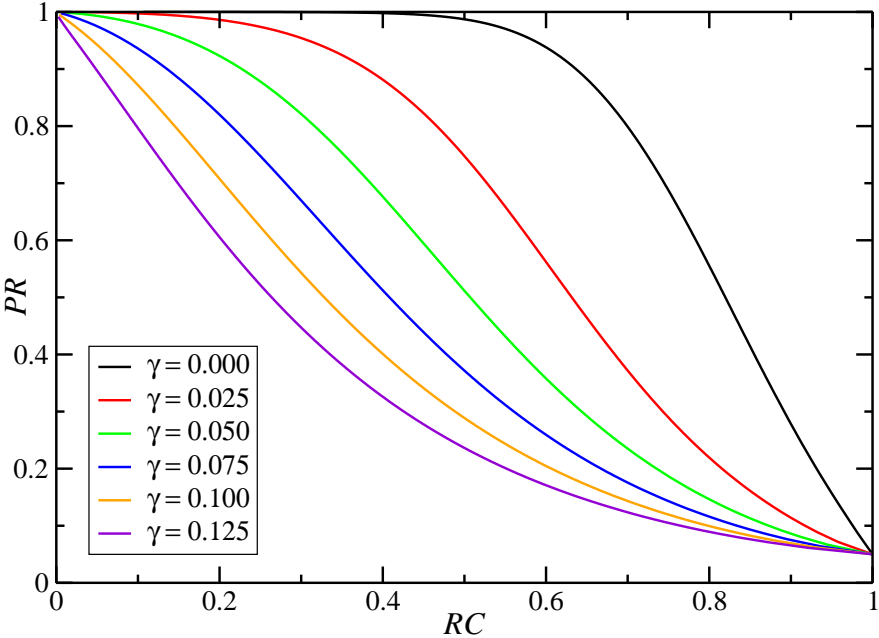}
\end{center}
\caption{Precision versus recall of BP for various noise strengths 
$\gamma = 0.0,\ 0.025,\ 0.05,\ 0.075,\ 0.1,\ 0.125$ (curves from right 
to left), ranging from no noise to equal signal and noise strengths.}
\label{fig:with_noise}
\end{figure}

\begin{figure}[htb]
\vspace{1.5cm}
\begin{center}
	\includegraphics[width=0.65\columnwidth]{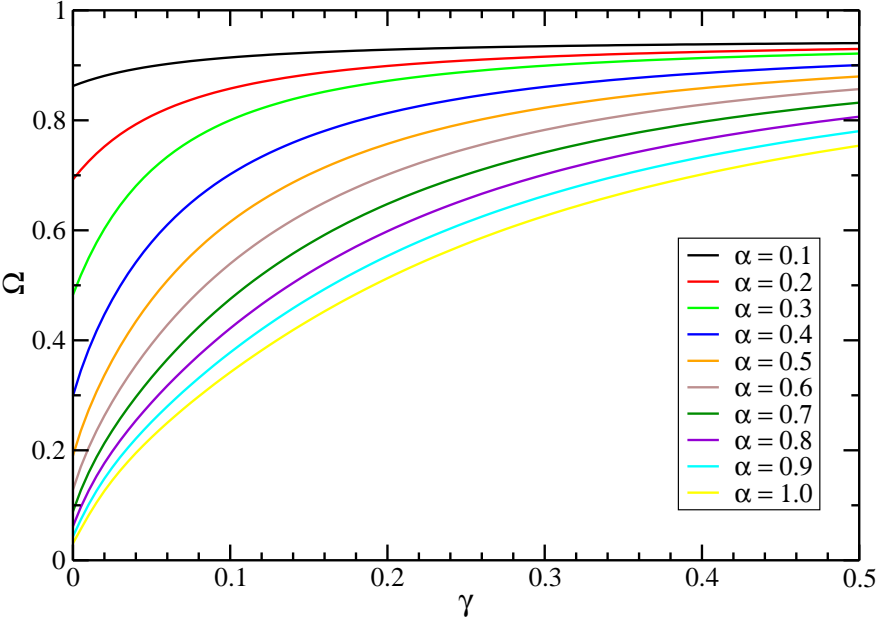}
\end{center}
\caption{Prediction error of BP as a function of the noise strength $\gamma$,
for various values of $\alpha$.}
\label{fig:prediction_error_gamma}
\end{figure}

To complete our analysis, we have also included errors into the data
generator, fixing all other parameters to $\alpha=0.5$ and 
$n_{eff}=0.04$. Other parameter values lead to qualitatively equivalent
results. In Figs.~\ref{fig:with_noise} and \ref{fig:prediction_error_gamma}
we have plotted the precision-vs.-recall curves and the prediction
error for various values $\gamma$ of the noise strength. The 
predictive power of the algorithms obviously decreases with increasing
noise level. Note, however, that even for $\gamma=k_1+4k_2$, where the
statistical properties of the signal and of the noise are equivalent,
a non-trivial prediction is possible.

\section{Conclusion and outlook}
In this paper, we have analysed the performance of two approaches to
the identification of gene-regulatory interactions. Whereas the first
method is based on a ranking of all gene pairs according to the
measured correlations between their expression levels, the second
method tries to include collective effects of activators and
repressors coming from the joint action of multiple regulators. This
second method, developed by the same authors in an earlier
publication, uses message-passing techniques (belief propagation)
which can be understood as an algorithmic reinterpretation of the
cavity method in spin-glass physics.

Using a simple data generator, we achieved a theoretical analysis of
the algorithmic performance. In the simpler case of pairwise
correlations, this analysis is based on a straight-forward application
of the central limit theorem. The more involved case of the second
algorithm requires the use of other spin-glass techniques, more
precisely of the replica method in order to perform a generalised
Gardner calculation in the space of all candidate gene networks. We
found that algorithms taking into account collective effects beyond
pair-correlations work better. The intuitive reason for this finding
is that the best joint prediction is not necessarily done by the set
of the best individual predictors. These results are true even if the
data generator is more complex than the inference algorithm. We tested
explicitly the cases of heterogeneous regulation strengths and noisy
data.

There are some interesting directions to be followed. First of all,
these algorithms are designed to be applied to real biological data.
As already seen in \cite{BP-infer}, also BP can be used to extract
sparse and predictive information from gene-expression data. Still
there is space for substantial improvements by integration of
biological knowledge which goes beyond a simple diluting field. One
possibility is, e.g., the integration of sequence information on
putative transcription factor binding sites via inhomogeneous diluting
fields.

One drawback in both the considered data generator and the derivation
of the BP algorithm is that we neglected correlations between
inputs. These correlations exist, however, in biological data, since
also transcription factors may be coregulated. An interesting approach
in this direction was recently presented by Kabashima \cite{Ka}, and
can be extended to our model. It would be interesting to see under
which conditions correlations really change the behaviour of the
algorithms.

A last point is the application of the algorithm to other problems.
In fact it gives more generally a sparse Bayesian
classifier. Bioinformatical applications of such classifiers goes far
beyond the inference of gene networks. One example we are currently
analysing is the classification of expression patterns of cancer
tissues according to their diagnosis (cancer vs. healthy tissues,
different cancer subtypes), and the detection of predictive but sparse
gene signatures for the different diagnoses \cite{UdKa,FraTrig}.

{\bf Acknowledgment:} We are grateful to Yoshiyuki Kabashima, Michele
Leone and Francesca Tria for interesting and helpful discussions.


\end{document}